\documentclass[prb,reprint,twocolumn,showpacs,superscriptaddress,floatfix]{revtex4-1}
\usepackage{graphicx,amssymb,amsmath,bbold,bm,wasysym,psfrag,color,tikz}
\graphicspath{{img/}}
\usepackage{mathtools}
\usepackage[normalem]{ulem}

\begin{document}
	
	\title{Dynamical and Topological Properties of the Kitaev Model in a $[111]$ Magnetic Field}
	\author{Matthias Gohlke}
	\affiliation{Max-Planck-Institut f\"ur Physik komplexer Systeme, 01187 Dresden, Germany}
	\author{Roderich Moessner}
	\affiliation{Max-Planck-Institut f\"ur Physik komplexer Systeme, 01187 Dresden, Germany}
	\author{Frank Pollmann}
	\affiliation{Technische Universit\"at M\"unchen, 85747 Garching, Germany}
	\date{\today}     
	
	\begin{abstract}
        The Kitaev model exhibits a quantum spin liquid
        hosting emergent fractionalized excitations. 
        We study the Kitaev model on the honeycomb lattice coupled to a magnetic field along the $[111]$ axis.
        Utilizing large scale matrix product based numerical models, we confirm three phases with
        transitions at different field strengths depending on the sign of the Kitaev
        exchange: a non-abelian topological phase at low fields,
        an enigmatic intermediate regime only present for antiferromagnetic Kitaev
        exchange, and a field-polarized phase.
        For the topological phase, we numerically observe the expected cubic scaling of the gap and extract the quantum dimension of the non-abelian anyons.
        Furthermore, we investigate dynamical signatures of the topological and the field-polarized phase using a matrix product operator based time evolution method.

	\end{abstract}
	\maketitle

	\section{Introduction} %

    Quantum spin liquids (QSLs) \cite{anderson_resonating_1973,savary_quantum_2017} are realized in certain spin systems where the interplay of frustration and quantum fluctuations suppresses long range order.
    These exotic phases of matter cannot be understood in terms of spontaneous symmetry breaking,
    but are instead characterized by their long range entanglement and emergent fractionalized excitations.
    The lack of local order parameters makes it difficult to experimentally detect QSLs by their static properties--except for showing the absence of conventional order.
    Instead it appears more promising to study dynamical properties of QSLs (e.g., the dynamical spin structure factor)
    which encode characteristic fingerprints of topological order\cite{qi_dynamics_2009,han_fractionalized_2012,dodds_quantum_2013,punk_topological_2014,morampudi_statistics_2017}.
   
    On the theory side, significant insight into the physics of QSLs comes from the study of exactly solvable models. 
    A prominent example is the Kitaev model on the honeycomb lattice\cite{kitaev_anyons_2006}, which exhibits a QSL phase
    featuring fractionalization of spin-$1/2$ degrees of freedom into fluxes and Majorana excitations.
    The Kitaev interaction, a strongly anisotropic Ising exchange appears to be realized approximately in compounds with strong spin-orbit interaction\cite{jackeli_mott_2009,witczak-krempa_correlated_2014,nussinov_compass_2015,rau_spin-orbit_2016,winter_models_2017},
    such as the iridates Na$_2$IrO$_3$, Li$_2$IrO$_3$~\cite{chaloupka_zigzag_2013}, and $\alpha$-RuCl$_3$~\cite{plumb_-rucl3:_2014,sears_magnetic_2015,banerjee_proximate_2016}.
    It may also be realized in metal-organic frameworks~\cite{yamada_designing_2017}.
    In such materials, additional interactions are important and typically lead to long-range magnetic order,
    nonetheless signatures of being in the proximity to the Kitaev QSL are discussed\cite{yadav_kitaev_2016,banerjee_proximate_2016,banerjee_neutron_2017}.
    Recent attention has shifted to applying a magnetic field\cite{yadav_kitaev_2016,janssen_honeycomb-lattice_2016,zheng_gapless_2017,janssen_magnetization_2017,jansa_observation_2017,sears_phase_2017,winter_probing_2018,gohlke_quantum_2018},
    in particular experiments on the Kitaev compound $\alpha$-RuCl$_3$ (with an in-plane magnetic field) reveal 
    a single transition into quantum paramagnetic phase with spin-excitation gap\cite{johnson_monoclinic_2015,ponomaryov_unconventional_2017,wolter_field-induced_2017,wang_magnetic_2017,banerjee_excitations_2018,lampen-kelley_anisotropic_2018,hentrich_unusual_2018}. 
    
    In this article, we consider the Kitaev model in a magnetic field along $[111]$,
    such that the field couples to the spins in a symmetry-equivalent way and the field does not prefer any bond in particular.
    While the magnetic field breaks integrability, Kitaev has identified two three-spin exchange terms
    within perturbation theory, that break time-reversal symmetry and open a gap in the spectrum.
    One of the terms retains integrability and upon adding to the Kitaev model,
    leads to a topologically ordered phase hosting non-abelian anyons\cite{kitaev_anyons_2006}.
    However, numerical simulations~\cite{jiang_possible_2011} 
    reveal that the same topological phase occurs for small magnetic fields and ferromagnetic Kitaev coupling.
    The topological phase turns out to be more stable, by one order of magnitude in the critical field strength,
    if an antiferromagnetic coupling is considered\cite{zhu_robust_2017}.
    Remarkably, an additional regime, possibly gapless, between the low-field topological and the high-field polarized phase appears to exist\cite{zhu_robust_2017}.
    
    In this work, we employ large scale infinite density matrix renormalisation group (iDMRG) methods\cite{white_density_1992,mcculloch_infinite_2008,phien_infinite_2012,kjall_phase_2013} to investigate the ground state phase diagram
    of the Kitaev model in a magnetic field along $[111]$ and simulate its dynamics using a matrix-product operator (MPO) based time-evolution\cite{zaletel_time-evolving_2015}.
    
    The topologically ordered phase is characterized by its finite topological entanglement entropy (TEE)\cite{levin_detecting_2006,kitaev_topological_2006}. 
    By subtracting contributions of the Majorana fermions and the $\mathbb Z_2$-gauge field
    from the numerically obtained entanglement entropy of a bipartition, 
    we extract a remainder which is identical to the TEE in the integrable case.
    In doing so, we obtain a clear signature of non-abelian anyonic quasiparticles in the topological phase.
    In a magnetic field, this remainder is still consistent with the existence of non-abelian anyons.
    
    Furthermore, within the topological phase the correlation length decreases with magnetic field in a way
    that is consistent with a cubic opening of the gap as found for the three-spin exchange\cite{kitaev_anyons_2006}.
    However, the dynamical spin-structure factor in presence of a field behaves very differently
    compared to what is known for the three-spin exchange\cite{knolle_dynamics_2015}.
    The magnetic field causes the flux degrees of freedom to become mobile.
    As a consequence the low-energy spectrum contains more structure and 
    the gap in the dynamical spin-structure factor is reduced.
    
    Approaching the intermediate regime from the polarized phase,
    the magnon modes reduce in frequency and simultaneously flatten.
    This resembles the phenomenology within linear spin wave theory (LSWT)\cite{janssen_honeycomb-lattice_2016,mcclarty_topological_2018},
    but the transition is significantly renormalised to lower fields.
    Close to the transition, 
    a broad continuum exists that, within our reachable resolution in frequency,
    reaches down to zero frequency and merges with the single magnon branches.
    At the transition, the spectrum appears to be (nearly) gapless in the entire reciprocal space.
    We do not observe an opening of a gap in the intermediate regime.

    The remainder of this paper is structured as follows:
    In Sec. \ref{scn:mod} we introduce the model consisting of Kitaev term,
    Zeeman coupling to a magnetic field along $[111]$,
    and three-spin exchange.
    In Sec. \ref{scn:gs}, the ground state phase diagram is discussed for both signs of the Kitaev coupling.
    We then focus on the antiferromagnetic Kitaev coupling in Sec. \ref{scn:dsf}
    and study its dynamical signatures within the low-field topological as well as the high-field polarized phases.
    We conclude with a summary and discussion in Sec.  \ref{scn:dis}.

    \section{Model} \label{scn:mod}

\begin{figure}[tb]
        \includegraphics{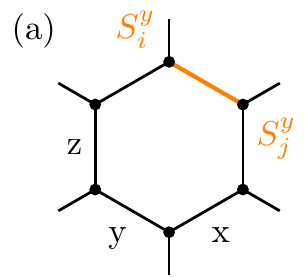}
        \includegraphics{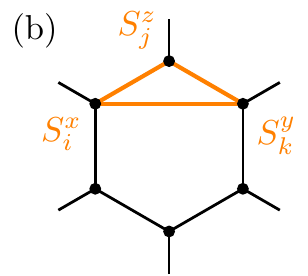}
        \caption{
            (a) Bonds labeled with x,y, and z and an exemplaric $S^y_i S^y_j$ Kitaev-exchange (orange),
            (b) a single three-spin term $S^x_i S^z_j S^y_k$ of the three-spin interaction in $H_{K_3}$.}
        \label{fig:H_and_K3}
    \end{figure}

    The Hamiltonian describing the Kitaev model in a magnetic field along [111] direction reads 
	\begin{equation}
		H =    \sum_{\langle i,j \rangle_\gamma} K_{\gamma} S_i^\gamma S_j^\gamma 
		   - h \sum_i \left( S^x_i + S^y_i + S^z_i \right) ~, 
		 \label{eqn:H_KH111}
	\end{equation}
    where the first term is the pure Kitaev model exhibiting strongly anisotropic spin exchange coupling\cite{kitaev_anyons_2006}. 
    Neighboring spins couple depending on the direction of their bond $\gamma$ with $S^x S^x$, $S^y S^y$ or $S^z S^z$, cf. Fig. \ref{fig:H_and_K3}(a).
    The second term is the Zeeman-coupling of the spins to a magnetic field applied in $[111]$ direction.
    
    In the zero field limit, the Kitaev model exhibits a quantum spin liquid ground state with fractionalized excitations\cite{kitaev_anyons_2006}.
    Depending on $K_\gamma$, the spectrum of the fermions is either \emph{gapped} (\emph{A-phase}) or \emph{gapless} (\emph{B-phase}).
    Let the $K_\gamma$ be sorted as $K_\alpha \ge K_\beta \ge K_\gamma$, then the gapless B-phase occurs if $|K_\alpha| \le |K_\beta| + |K_\gamma|$
    and the A-phase if $|K_\alpha| > |K_\beta| + |K_\gamma|$.
    In the remainder, we consider the isotropic case $K_\gamma = K = \pm 1$.
    
    Flux degrees are defined by the plaquette operator $W_p = \prod_{i \in \mathcal P} \sigma^{\gamma(i)}_i$,
    where $\gamma(i) = \{x,y,z\}$ equals the bond, that is not part of the loop $\mathcal P$ around the plaquette.
    The $W_p$ commute with the Hamiltonian (in the $h=0$ limit) and have eigenvalues $\pm 1$.
    Thus, the $W_p$'s are quantum numbers separating the full Hilbert space into subspaces,
    for each of which a free fermion problem remains to be solved.
    The ground state lies in the flux-free sector, that is $\forall i: W_{p,i} = +1$.

    For later use, we comment on placing the Kitaev model on a cylinder.
    A second flux operator of a non-contractable loop $\mathcal C$ going around the cylinder can be defined:
    $W_l = \prod_{i\in \mathcal C} \sigma^{\gamma(i)}_i$.
    Similarly to $W_p$, $W_l$ commutes with the Hamiltonian, has eigenvalues $\pm 1$, and separates the full Hilbert space in two subspaces.
    With respect to the free fermions, $W_l = -1$ (flux-free) corresponds to periodic
    and $+1$ to antiperiodic boundary conditions along the circumference of the cylinder.
    The ground state within each of the two sectors are separated in energy by $\Delta E$,
    which depends on the circumference $L_\text{circ}$ and vanishes in the limit $L_\text{circ} \rightarrow \infty$.
    
    Applying a magnetic field $h$ along $[111]$, as in Eq. (\ref{eqn:H_KH111}), breaks time-reversal symmetry
    and opens a gap in the fermionic spectrum.
    The lowest order terms breaking time-reversal and not changing the flux configuration
    are the three-spin exchanges $S_i^x S_j^y S_k^z$.
    Two such terms exist\cite{kitaev_anyons_2006}.
    The one illustrated in Fig.~\ref{fig:H_and_K3}(b) plus symmetric variants
    results in a quadratic Hamiltonian for the Majorana fermions 
    and thus preserves the integrability of the original model.
    The corresponding Hamiltonian reads
	\begin{equation}
		H_{K_3} =  \sum_{\langle i,j \rangle_\gamma} K_{\gamma} S_i^\gamma S_j^\gamma 
		   + K_3 \sum_{\langle \langle i,j,k \rangle \rangle} S_i^x S_j^y S_k^z ~, 
		 \label{eqn:H_K3}
	\end{equation}
    where $\langle \langle . \rangle \rangle$ denotes an ordered tuple $(i,j,k)$ of neighboring sites such that the $S^x$, $S^y$, and $S^z$ at the outer two sites coincide with the label of the bond connecting to the central site.
    The flux operators $W_p$ and $W_l$ still commute with $H_{K_3}$ and separate the Hilbert space.
    The remaining fermionic Hamiltonian is quadratic with the corresponding bands having non-zero Chern number $\pm 1$ and
    yielding composite excitations with anyonic exchange statistics\cite{kitaev_anyons_2006}.
    
    \section{Ground State Phases} \label{scn:gs} 
    
    \begin{figure}
        \centering
        \includegraphics[]{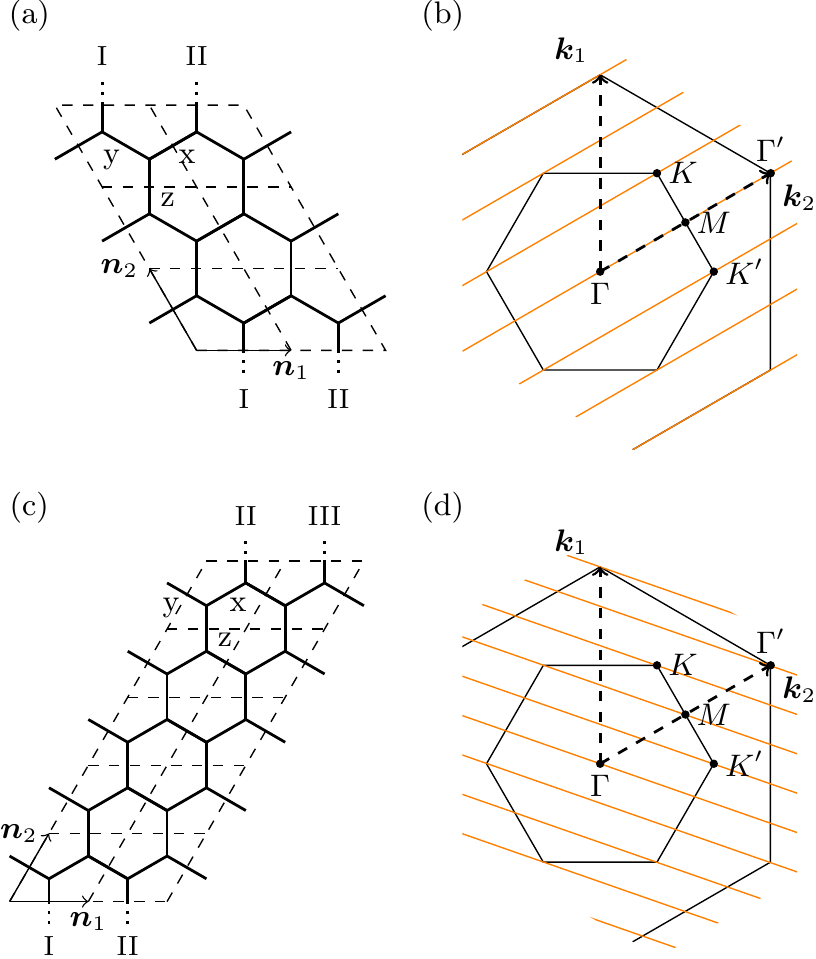}
        \caption{
            Geometries used for iDMRG and their corresponding accessible momenta (orange lines) in reciprocal space with respect to the first Brillouin zone (inner hexagon).
            The second Brillouin zone is shown partially.
            The roman numbers label links across the boundary.
            (a) \emph{rhombic} geometry with three unit cells, $L_\text{circ} = 6$ sites, along the circumference
            and (b) its corresponding reciprocal space.
            (c,d) \emph{rhombic-2} geometry with five unit cells circumference, $L_\text{circ} = 10$ sites.
            }
        \label{fig:lat_BC}
    \end{figure}
	
    The ground state is obtained using the \emph{matrix product state} (MPS) based \emph{infinite density matrix renormalisation group} (iDMRG) method~\cite{white_density_1992,mcculloch_infinite_2008,phien_infinite_2012,kjall_phase_2013}.
	Being a standard technique for one-dimensional systems, 
    it has been used in two dimensions by wrapping the lattice on a cylinder
    and mapping the cylinder to a chain with longer range interactions.
	
    We employ a \emph{rhombic-2} geometry with a circumference of $L_\text{circ} = 10$ sites
    and a \emph{rhombic} geometry with $L_\text{circ}=6$ as illustrated in Fig.~\ref{fig:lat_BC}.
	Both geometries capture the $K-$points in reciprocal space and hence are gapless for pure Kitaev-coupling ($h = 0$).
	A main advantage of the \emph{rhombic-2} geometry is its translational invariance of the chain winding around the cylinder.
    While the mapping to a cylinder for the \emph{rhombic} geometry 
    requires an iDMRG unit cell of at least $L_\text{circ}$ sites, 
    a single fundamental unit cell with two sites is sufficient to simulate an infinite cylinder
    using the \emph{rhombic-2} geometry.
    Different iDMRG cells have been used to test for possible breaking of translational symmetry
    and corresponding results will be presented when of relevance.
	We use bond dimensions of up to $\chi = 1600$ for the computation of the phase diagram.

    \begin{figure}
		\includegraphics[width=\linewidth]{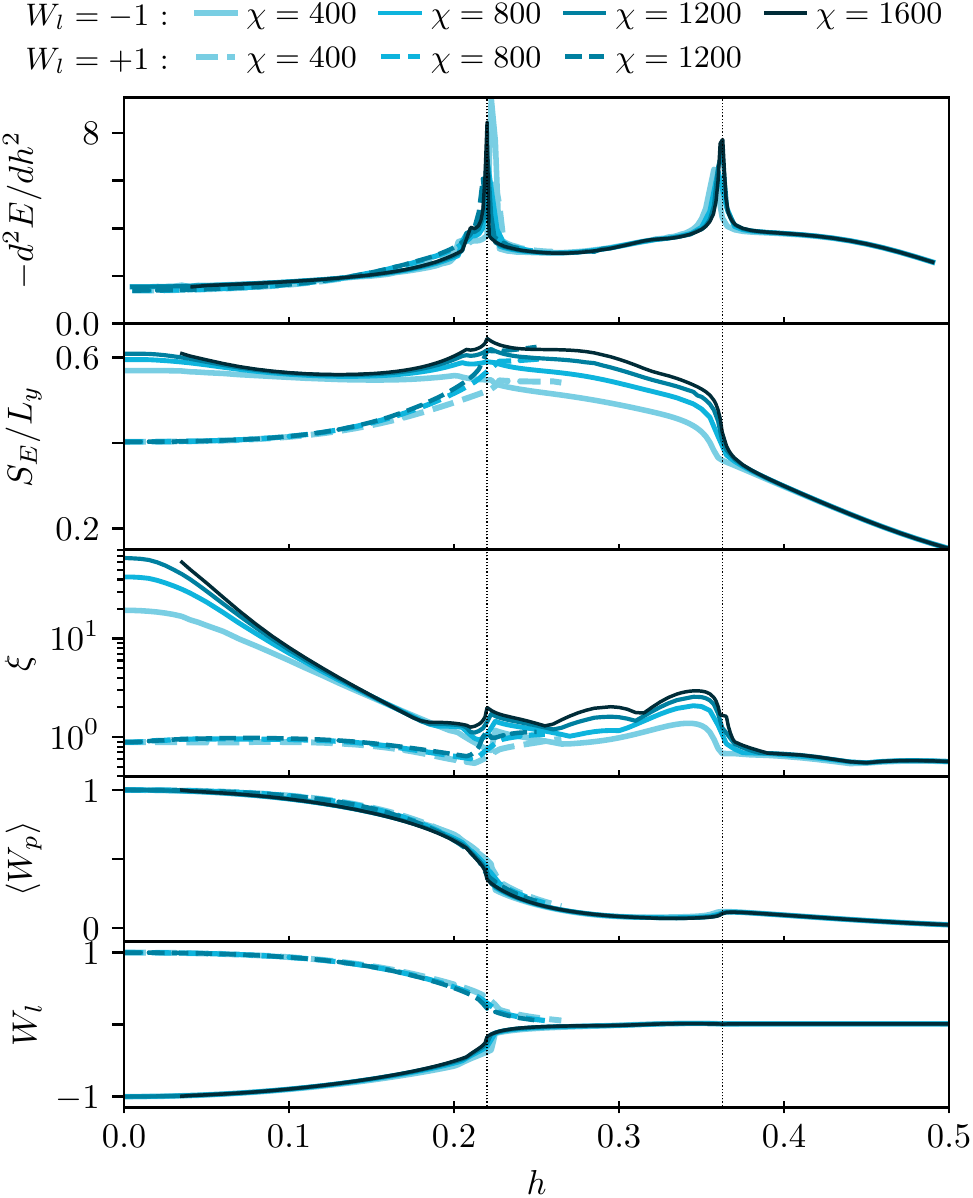}
		\caption{
            Several observables of the Kitaev model with antiferromagnetic coupling, $K > 0$, in a magnetic field along $[111]$.
            From top to bottom:
            Second derivative $-d^2 E/d h^2$ of the energy with respect to the field $h$,
            entanglement entropy $S_E$ of a bipartition of the cylinder divided by the number $L_y$ of bonds cut,
            correlation length $\xi$,
            average of plaquette fluxes $W_p$,
            and flux $W_l$ of a non-contractible loop around the cylinder.
            At least three phases are observed: topological phase for $h < h_{c1,AF} \approx 0.22$,
            intermediate possibly gapless $h_{c1,AF} < h < h_{c2,AF}\approx0.36$, and a subsequent field-polarised phase.
            Solid blue lines are for the $W_l=-1$ sector, dashed blue lines for $W_l=+1$,
            and its intensity encodes the bond dimension $\chi$ used, where dark blue refers to a large $\chi$.
            Thin dashed black lines depict the phase transitions obtained from the peaks in $-d^2 E/d h^2$.
        }
        \label{fig:pd}
	\end{figure}
	
    We confirm the existence of two phases and a single transition for ferromagnetic Kitaev coupling\cite{jiang_possible_2011, zhu_robust_2017} (FMK, $K < 0$), 
    and of at least three phases for antiferromagnetic Kitaev coupling\cite{zhu_robust_2017} (AFK, $K>0$).
    For both, FMK and AFK, we find a topological phase at low field and a field-polarized phase at high field.
    Only for AFK, we identify an intermediate, seemingly gapless, phase.
	
    \subsection{Topological Phase} \label{scn:gs_topo}
	
    For small $h$, the system forms a non-abelian topological phase\cite{kitaev_anyons_2006}.
	Its stability upon applying $h$ vastly differs depending on the sign of the Kitaev interaction. 
	Employing a \emph{rhombic-2} geometry with $L_\text{circ}=10$, we find, in case of AFK,
    that this phase ranges up to $h_{c1,AF} \approx 0.22$, whereas for FMK it ranges only up to $h_{c,FM} \approx 0.014$.
    Both values are based on the peaks in the second derivative $-d^2E/dh^2$ of the energy with respect to the magnetic field.
    However, subtle features are present for AFK at slightly lower $h \approx 0.2$,
    which become less pronounced with larger bond dimension $\chi$. 
    In comparison to values reported earlier\cite{jiang_possible_2011, zhu_robust_2017} we find a nearly $30\%$ lower value for the FMK transition $h_{c,FM}$. 
    This is due to the fact that for rather small circumferences,
    the ground state energy within the topological phase is strongly sensitive to the boundary condition
    as has already been noted in Ref.~[\onlinecite{kitaev_anyons_2006}].
    The {\emph{rhombic-2}} geometry we employ has the same twisted boundary condition
    as the $(L \bm n_1, L \bm n_2 + \bm n_1)$ geometry employed in [\onlinecite{kitaev_anyons_2006}],
    which is shown to converge better in energy when increasing $L$ or $L_\text{circ}$, respectively.
    The transition field $h_{c,FM}$ may still decrease slightly upon further increasing $L_\text{circ}$
    and approaching the two-dimensional limit $L_\text{circ} \rightarrow \infty$. 

    For small $h$, the total magnetisation, $|\langle \bm S \rangle|$ (not shown here), grows proportionally with $h$.
    The two sectors found on the cylindrical geometry and determined by $W_l = +1$ or $W_l = -1$ are distinguished
    by their behaviour of the entanglement entropy $S_E$ and the correlation length $\xi$.
    The $W_l = +1$ sector is characterized by finite $\xi$ and $S_E$
    due to being gapped by imposing antiperiodic boundary conditions on the Majorana fermions\cite{gohlke_dynamics_2017}.
    In contrast, the $W_l = -1$ sector has divergent $\xi$ and $S_E$ when $h=0$, where it is gapless.
    In the latter, encoding the wave function as MPS with a finite $\chi$ induces an effective gap that limits $\xi$ and $S_E$. 
    In fact, the growth of $\xi$ and $S_E$ with increasing $\chi$ is connected via
    $S_{E,\chi} = c/6 \log \xi_\chi + \text{const}$~\cite{calabrese_entanglement_2004,tagliacozzo_scaling_2008},
    where $c$ is the universal \emph{central charge}.
    This has been named finite entanglement scaling and allows to confirm $c=1$ (for $h=0$, $W_l=-1$)
    as has been checked previously on a different cylinder geometry\cite{gohlke_dynamics_2017}.
    As a side remark, the notion of a central charge is applicable due to using a cylinder geometry
    and effectively mapping the model in question to a one-dimensional system.

    In a magnetic field, $\langle W_p \rangle$ as well as the cylinder flux $W_l$ begin to slowly deviate from $\pm 1$
    until they vanish close to the transition.
    The plaquette fluxes $W_p$, as defined in the integrable limit, are not conserved anymore for finite $h$
    as the application of a single $S^\gamma_i$ creates a flux each on the two plaquettes adjacent to bond $\gamma$ at site $i$.
    However, an adiabatically connected operator $\tilde W_l$ of $W_l$ is expected to exist,
    such that $\tilde W_l \approx \pm1$~\cite{hastings_quasiadiabatic_2005}.
    Such a dressed Wilson loop $\tilde W_l$ separates the two sectors found on the cylinder
    for any $h$ within the topological phase.
    
    \begin{figure}
		\includegraphics[width=\linewidth]{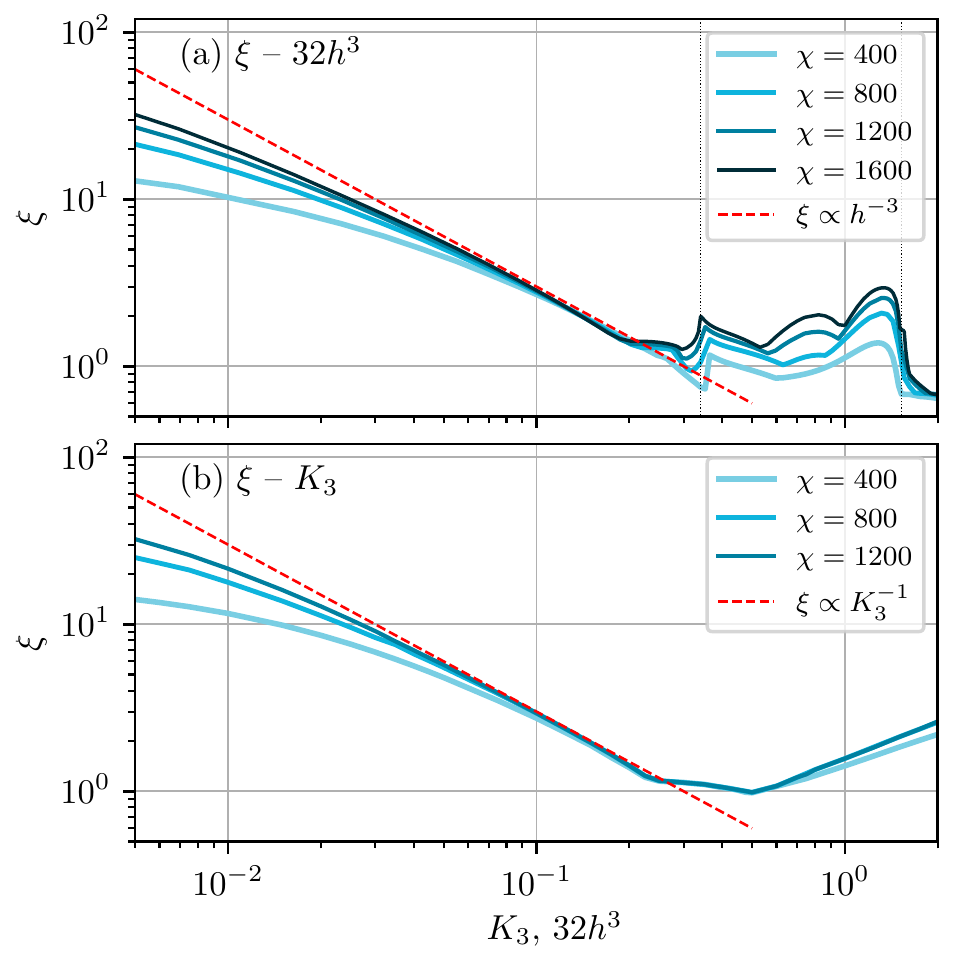}
		\caption{
            Comparison of correlation length $\xi$ between
            (a) the rescaled magnetic field $h \rightarrow 32 h^3$ and
            (b) the three-spin exchange $K_3$.
            The solid red line is a guide-to-the-eye corresponding to a $1/K_3$ or $1/(32h^3)$ scaling.
            Within $0.05 < K_3, 32h^3 < 0.2$, that is where numerical convergence is achieved,
            the behaviour of $\xi$ is consistent with a predicted opening of the gap as $\Delta E \propto h^3$ or as $\Delta E \propto K_3$, respectively.
        }
        \label{fig:top_hp3_K3}
    \end{figure}
    
    Numerical convergence, that is $\xi$ and $S_E$ become $\chi$-independent, is achieved for $0.1 < h < 0.18$.
    In that range $\xi$ reflects the physical excitation gap\cite{hastings_locality_2004} via $\Delta E \propto 1/\xi$.

    Figure \ref{fig:top_hp3_K3}, where the x-axis has been rescaled $h \to 32h^3$,
    enables a direct comparison with the three-spin exchange $K_3$ in $H_{K_3}$.
    Both exhibit a very similar decrease of $\xi$ with a $\xi \propto 1/x$ scaling, where $x$ is either $32h^3$ or $K_3$.
    $\xi$ reaches a plateau at $x=0.2$ with a low $\xi \approx 1$.
    If $h$ is applied, a small $\chi$-dependent dip and the phase transition into the intermediate regime follows,
    whereas for $K_3$ the plateau ranges up to $K_3 = 1$, from where $\xi$ increases again%
    \footnote{For large $K_3 \gg 1$, the flux gap reduces and vanishes.
    The ground state is then not in a flux-free sector anymore.}.
    The entanglement entropy $S_E$ reaches, in the case of a magnetic field,
    a plateau already at $32h^3 \approx 0.06$ ($h \approx 0.12$)
    beyond which it raises again until the transition field $h_{c1,AF}$ is reached.
    At all fields the entanglement remains larger than for the corresponding $K_3$.
    A more detailed discussion about the entanglement in the context of topological excitations and topological entanglement entropy follows below. 
    The $W_l=+1$ sector has $\chi$-independent $\xi$ and $S_E$ up to $h \approx 0.18$.
    Before the transition ($0.18 < h < h_{c1,AF}$) both sectors exhibit a $\chi$-dependents
    which suggests a closing of the gap at the transition 
    and, thence, indicates that the transition might be continuous.
    
	\begin{figure}[tb]
		\includegraphics[width=\linewidth]{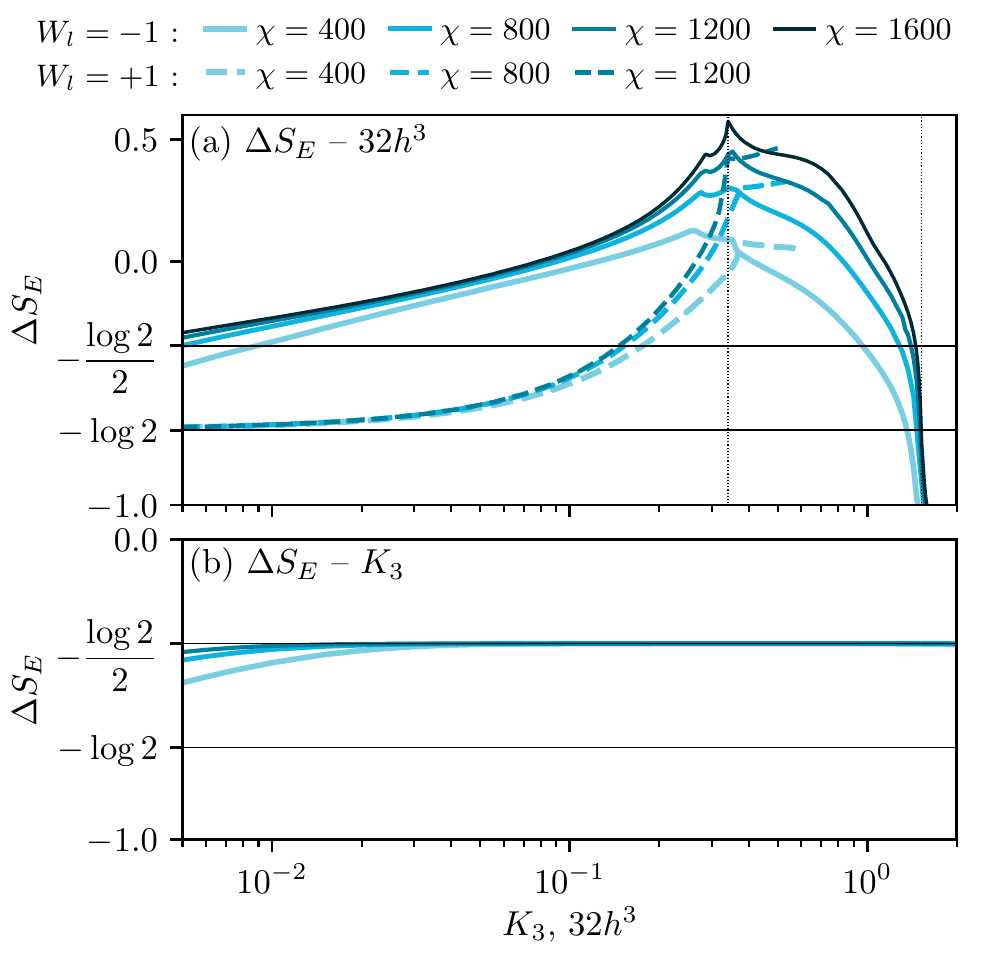}
		\caption{
            Remainder $\Delta S_E$ of the entanglement entropy of a bipartition of the cylinder after subtracting
            a fermionic and a gauge field contribution following Eq. (\ref{eqn:DS_E}).
            The magnetic field has been rescaled, $h\rightarrow 32h^3$, based on the behaviour of the correlation length in Fig. \ref{fig:top_hp3_K3}.
            The vertical dashed lines signal the transitions in a magnetic field.
            The horizontal lines correspond to $\log (\mathcal D/d_a)$ as discussed in the main text.
            }
        \label{fig:topSbip}
	\end{figure}

    We now focus on the characterization of the topological order occurring at low magnetic fields $h$ or when non-zero $K_3$ is considered.
    First, let us recall some facts about topologically ordered systems on an infinite cylinder\cite{cincio_characterizing_2013,zaletel_topological_2013}.
    Generally, topological order leads to a ground state degeneracy
    with a number of degenerate states being equal to the number of emergent quasiparticle species.
    These ground states are conveniently represented as minimally entangled states (MES)\cite{zhang_quasiparticle_2012,zaletel_topological_2013}, say $|\psi_0^i\rangle$,
    where $i$ denotes the particular quasiparticle.
    By utilizing iDMRG, such MES are selected naturally,
    and the obtained MPS corresponds to one of the quasiparticles\cite{jiang_identifying_2012,cincio_characterizing_2013}. 
    
    Upon cutting a cylinder into two semi-infinite halves, the entanglement entropy grows proportional with the circumference $L_\text{circ}$ as\cite{kitaev_topological_2006}
    \begin{equation}
        S_E = \alpha L_\text{circ} - \gamma_i~,
        \label{eqn:S_E_area_law}
    \end{equation}
    where $\gamma_i$ denotes the topological entanglement entropy (TEE)\cite{levin_detecting_2006,kitaev_topological_2006}.
    A non-zero TEE $\gamma_i = \log(\mathcal D/d_i)$ reveals topological order and is connected to the total quantum dimension $\mathcal D$,
    which itself is a sum of the quantum dimension $d_i$ of each quasiparticle
    \begin{equation}
        \mathcal D = \sqrt{\sum_i d_i^2}~.
        \label{eqn:totalQD}
    \end{equation}
    The quantum dimension is related to the fusion vector space,
    which is spanned by all the different ways anyons can fuse to yield a trivial total charge\cite{kitaev_topological_2006,nayak_non-abelian_2008}.
    The quantum dimension of abelian anyons is $d_i=1$, whereas for non-abelian anyons $d_i$ is generally larger than one.
    The gapped phase of the Kitaev model upon applying $K_3$ is known to exhibit topological order hosting non-abelian Ising anyons\cite{kitaev_anyons_2006}.
    The following quasiparticles exist: $\mathbb{1}$ (vacuum), $\epsilon$ (fermion), and $\sigma$ (vortex),
    of which $\sigma$ has a quantum dimension $d_\sigma=\sqrt 2$ and the other two $d_\mathbb{1} = d_\epsilon =1$. 
    From (\ref{eqn:totalQD}) follows a total quantum dimension of $\mathcal D = 2$.

    The Kitaev model has two separate contributions\cite{yao_entanglement_2010} to the entanglement entropy
    \begin{equation}
        S_E = S_G + S_F~.
        \label{eqn:S_G_+_S_F}
    \end{equation}
    The first contribution, $S_G$, originates from the static $\mathbb Z_2$-gauge field and is stated to be\cite{yao_entanglement_2010,dora_gauge_2018}
    \begin{equation}
        S_G = \left(\frac{N_y}{2}-1\right) \log 2~,
        \label{eqn:S_G}
    \end{equation}
    where $N_y$ is the number of unit cells along the circumference
    and equals the number of bonds cut by the bipartition, thus $N_y=L_\text{circ}/2$.
    The second contribution, $S_F$, describes the entanglement of the matter fermions\cite{yao_entanglement_2010}.
    By comparison with Eq.~(\ref{eqn:S_E_area_law}), the constant part in~(\ref{eqn:S_G}) resembles the TEE $\gamma_i = \log 2$.

    We turn to our iDMRG results now,
    where the entanglement entropy is readily available from the MPS representation of the ground state wave function.
    As will become clear later, we consider the following quantity
    \begin{equation}
        \Delta S_E = S_E - S_F - \frac{N_y}{2} \log 2 ~\approx \gamma_i ~,
        \label{eqn:DS_E}
    \end{equation}
    where $S_E$ is the entanglement entropy extracted numerically using iDMRG.
    $S_F$ can be computed exactly via the eigenvectors of the fermion hopping matrix
    if $H_{K_3}$ is considered\cite{yao_entanglement_2010,meichanetzidis_anatomy_2016}.
    We compute $S_F$ on a torus with one dimension equalling $L_\text{circ}$ and the second dimension being much larger.
    Note that a bipartition of a torus leaves two cuts of length $L_\text{circ}$,
    whereas on the cylinder there is only one such cut.
    Thence, only half of $\tilde S_F$ of a torus is considered in Eq.~(\ref{eqn:DS_E}).

    In the exactly solvable case of $H_{K_3}$, $\Delta S_E$ reproduces the TEE,
    such that $\Delta S_{E,K_3} = \gamma_i$ for all $K_3$,
    except when iDMRG is not converged with respect to $\chi$.
    From Fig.~\ref{fig:topSbip}, we recover the following TEE
    \begin{equation}
        \gamma_i =
            \begin{cases}
                \log 2 \quad & (W_l=+1), \\
                \log \frac{2}{\sqrt 2} \quad & (W_l=-1), 
            \end{cases}
        \label{eqn:iDMRG_gamma}
    \end{equation}
    which depends on the sector $W_l=\pm1$.
    In the gapless limit of the $W_l=-1$ sector ($K_3 = 0$),
    $S_F$ is divergent.
    Thus, at small $K_3$ the MPS improves with increasing $\chi$ similar to the behaviour of $\xi$ discussed before.
    Nonetheless, from (\ref{eqn:iDMRG_gamma}) a total quantum dimension of $\mathcal D = 2$ can simply be read off.
    The $W_l=-1$ sector contains a non-abelian anyon, a vortex $\sigma$, with quantum dimension $d_i=\sqrt 2$. 
    The ground state of the $W_l=+1$ sector is doubly degenerate with $d_i = 1$ for both states.
    Thus, the expected degeneracy is recovered.
    
    Upon applying the magnetic field, the integrability of $H_{K_3}$ in Eq.~(\ref{eqn:H_K3}) is lost
    and the fermionic contribution $S_F$ cannot be computed exactly. 
    Based on the fact that we observe a similar opening of the gap in the fermionic spectrum for $K_3$ and $h$
    when the magnetic field is rescaled as $h \rightarrow 32h^3$, 
    we assume that $S_F$ as a function of the rescaled magnetic field $S_F(32h^3)$
    is similar to $S_F(K_3)$ as a function of $K_3$.
    This assumption is at least justified in the limit of small $h$. 
    Figure~\ref{fig:topSbip}(a) displays $\Delta S_E$ in a magnetic field,
    where it approaches the same values of $\gamma_i$ for small $h$. 
    At elevated fields, $\Delta S_E$ begins to deviate from $\gamma = \log 2$ or $\gamma_\sigma = \log \sqrt 2$.
    $\Delta S_E$ increases monotonically until the transition into the intermediate phase is reached.

    In a magnetic field, the separability of fluxes and fermions is lost
    and generically additional entanglement between fluxes and fermions is created.
    Such entanglement generates an additional contribution $S_{F \otimes G}$ to the entanglement entropy,
    which is not accounted for in Eqs.~(\ref{eqn:S_G_+_S_F}) and (\ref{eqn:DS_E}).
    As this deviation occurs continuously, we like to argue
    that the topological phase in a low magnetic field is adiabatically connected to the topological phase of $H_{K_3}$ at non-zero $K_3$.

    As a remark, the difference of $\Delta S_E$ between the $W_l=\pm 1$ sectors is not constant.
    This is due to the correlation length of the fermions being enhanced in the $-1$ sector,
    particularly near the gapless limit ($h=0$), where it diverges. 
    Thus, the fermions may build up entanglement with the fluxes in an increased area near the cut resulting in an enhanced $S_{F\otimes G}$.

    We like to conclude that we find numerical evidence for a total quantum dimension $\mathcal D = 2$
    with non-abelian anyons having quantum dimension $d_i = \sqrt 2$ in the exactly solvable model using the three-spin term.
    The results using the magnetic field, breaking integrability of the original model, are still consistent with the results above.    However, a significant contribution to the entanglement entropy arises at increased magnetic fields. 

	\subsection{Intermediate Regime}
    \label{sec:itmd}

    Only for AFK, an intermediate region exists ranging from $h_{c1,AF} < h < h_{c2,AF}$,
    where $h_{c1,AF}\approx0.22$ (for \emph{rhombic-2}, $L_\text{circ}=10$) marks the transition from the topological phase
    and $h_{c2,AF}\approx0.36$ the transition into the field-polarised phase.
    
    The ground state within the intermediate regime requires to go to comparably large bond dimensions $\chi~\approx~1000$.
    Using smaller $\chi$, the ground state is very sensitive to the cylindrical geometry
    as well as the size of the iDMRG cell.
    However, based on the $1/\chi$-extrapolation of the ground state energy, that is presented in Appendix~\ref{app:fs_int},
    we find evidence for a translationally invariant ground state.
    In particular, when using a larger iDMRG cell,
    we observe a restoration of translational symmetry upon reaching a sufficiently large $\chi$.

    This motivates the use of the \emph{rhombic-2} geometry with an iDMRG cell equivalent to a single fundamental unit cell,
    which on the one hand suppresses ground states with enlarged unit cells due to broken translational symmetry,
    but on the other hand saves computational resources better spent in reaching larger $\chi$.

    Returning to its physical properties, the intermediate region exhibits a behaviour typical for a gapless phase.
    Both correlation length $\xi$ and entanglement entropy $S_E$ are not converged with respect to $\chi$,
    where $\xi$ increases slowly with $\chi$, while $S_E$ increases somewhat faster than in the gapless Kitaev limit.
    As we are studying effectively a one-dimensional system due to the cylindrical geometry,
    the finite-$\chi$ scaling\cite{tagliacozzo_scaling_2008} extracting a central charge may be applicable\cite{geraedts_half-filled_2016}.
    In that context, the behaviour of $S_E$ and $\xi$ indicate a larger central charge $c$,
    than found in the B-phase of the bare Kitaev model.
    However, the finite-$\chi$ scaling, see also Appendix \ref{app:fs_int}, does not reveal a conclusive $c$.
    Furthermore, the behaviour $\xi$ for larger $\chi \ge 800$ suggests a separation of the intermediate region into three phases,
    of which the middle one grows in extent with larger $\chi$.
    Given the large entanglement and the sensitivity to boundary conditions,
    our iDMRG results can only be suggestive for the nature of the ground state in the two-dimensional limit. 

    The flux expectation values $W_p$ and $W_l$ approach zero continuously.
    Interestingly, the coexistence of both sectors found in the topological phase, $W_l|_{h=0} = \pm1$,
    persists beyond the transition $h_{c1,AF}$.
    The peak in $-d^2E/dh^2$ signaling this transition is independent of the particular sector.

	\subsection{Polarized Phase}
    A transition to the large-$h$ field-polarized phase occurs at $h_{c2,AF}\approx0.36$ (AFK),
    or $h_{c,FM}\approx0.014$ (FMK), respectively.
    The polarized phase is gapped,
    which is signaled by the DMRG simulations by a finite correlation length $\xi$ and finite entanglement entropy $S_E$.
    The entanglement $S_E$ decreases with increasing field $h$ and vanishes once the magnetic moments approach saturation,
    where the ground state is a simple product state.
    At the transition both, FMK and AFK,
    exhibit a longitudinal magnetic moment of $\approx 55\%$ of saturation along the $[111]$ direction
    without any transverse component.
    The longitudinal moment grows with $h$ reaching $90\%$ saturation near $h\approx0.6$ (AFK) and $h\approx0.2$ (FMK).
    Large magnetic moments motivate perturbative methods like spin wave-theory\cite{mcclarty_topological_2018}.
    In comparison to linear spin wave theory (LSWT)\cite{janssen_honeycomb-lattice_2016},
    the transition gets renormalized significantly from $h_{LSW,AF}=1/\sqrt 3\approx 0.58$ down to $h_{c2,AF}$.
    For FMK, LSWT predicts a transition at exactly zero\cite{janssen_honeycomb-lattice_2016},
    whereas in iDMRG it occurs at small, non-zero field.

    \section{Dynamical Spin-Structure Factor} \label{scn:dsf}
    
    \begin{figure*}
        \includegraphics[width=\linewidth]{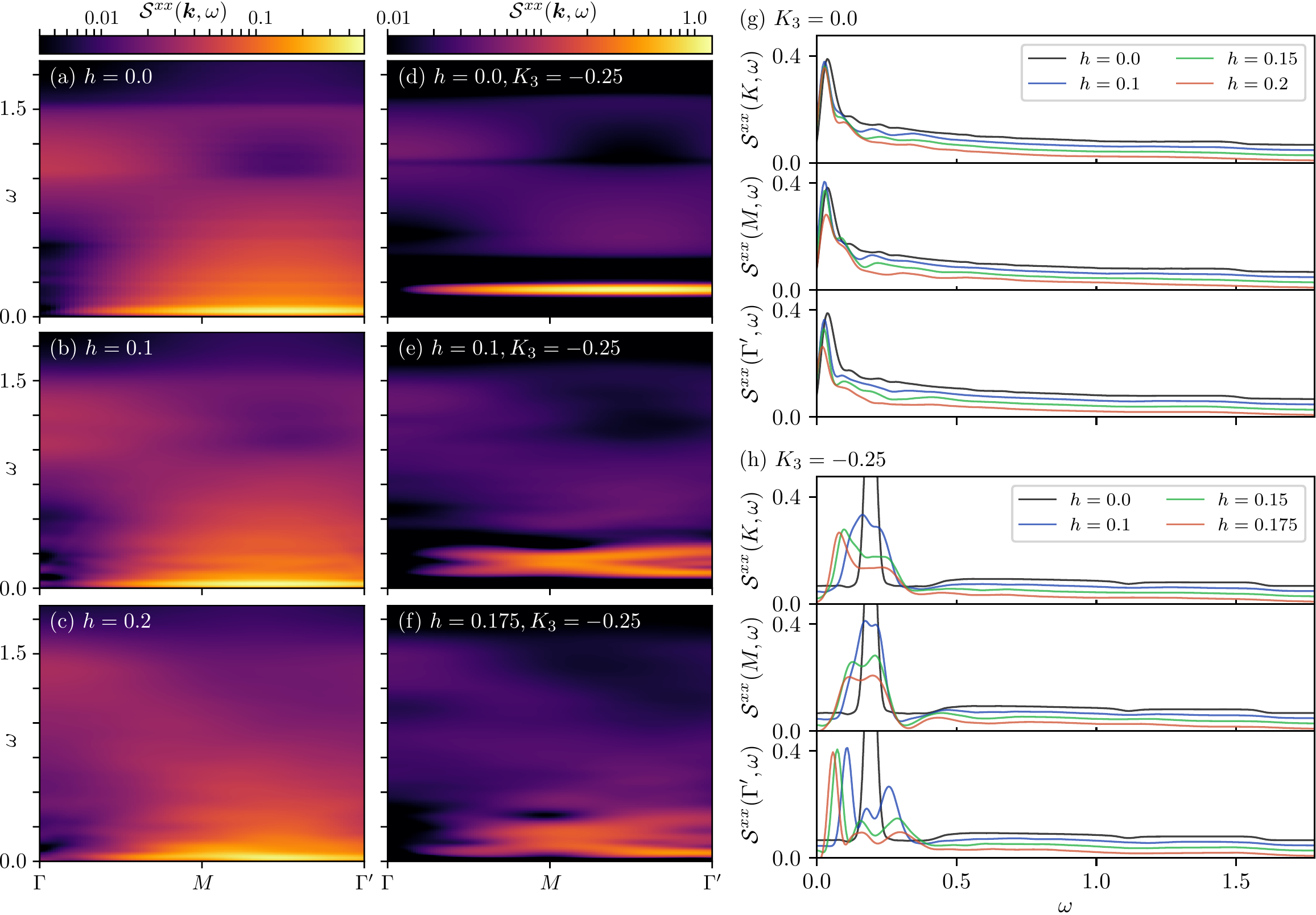}
        \caption{
            Dynamical spin-structure factor $\mathcal S^{xx}(k,\omega)$ along $\Gamma$--$M$--$\Gamma'$ at: 
            (a) $h=0.0$, (b) $h=0.1$, (c) $h=0.2$,
            (d) $h=0.0, K_3=-0.25$, (e) $h=0.1, K_3=-0.25$, (f) $h=0.175, K_3=-0.25$
            within the topological phase.
            (a)-(f) have a logarithmic color scale ranging over two decades. 
            (g,h) $\mathcal S^{xx}(k,\omega)$ at high-symmetry points $\Gamma$, $K$, $M$, and $\Gamma'$
            for different $h$ and $K_3$.
            An vertical offset is used for better visibility.
            In all plots, $\mathcal S^{xx}(k,\omega)$ is normalized as given in the main text.
        }
        \label{fig:dsf}
    \end{figure*}
 
    \begin{figure*}[tb]
         \includegraphics[width=\linewidth]{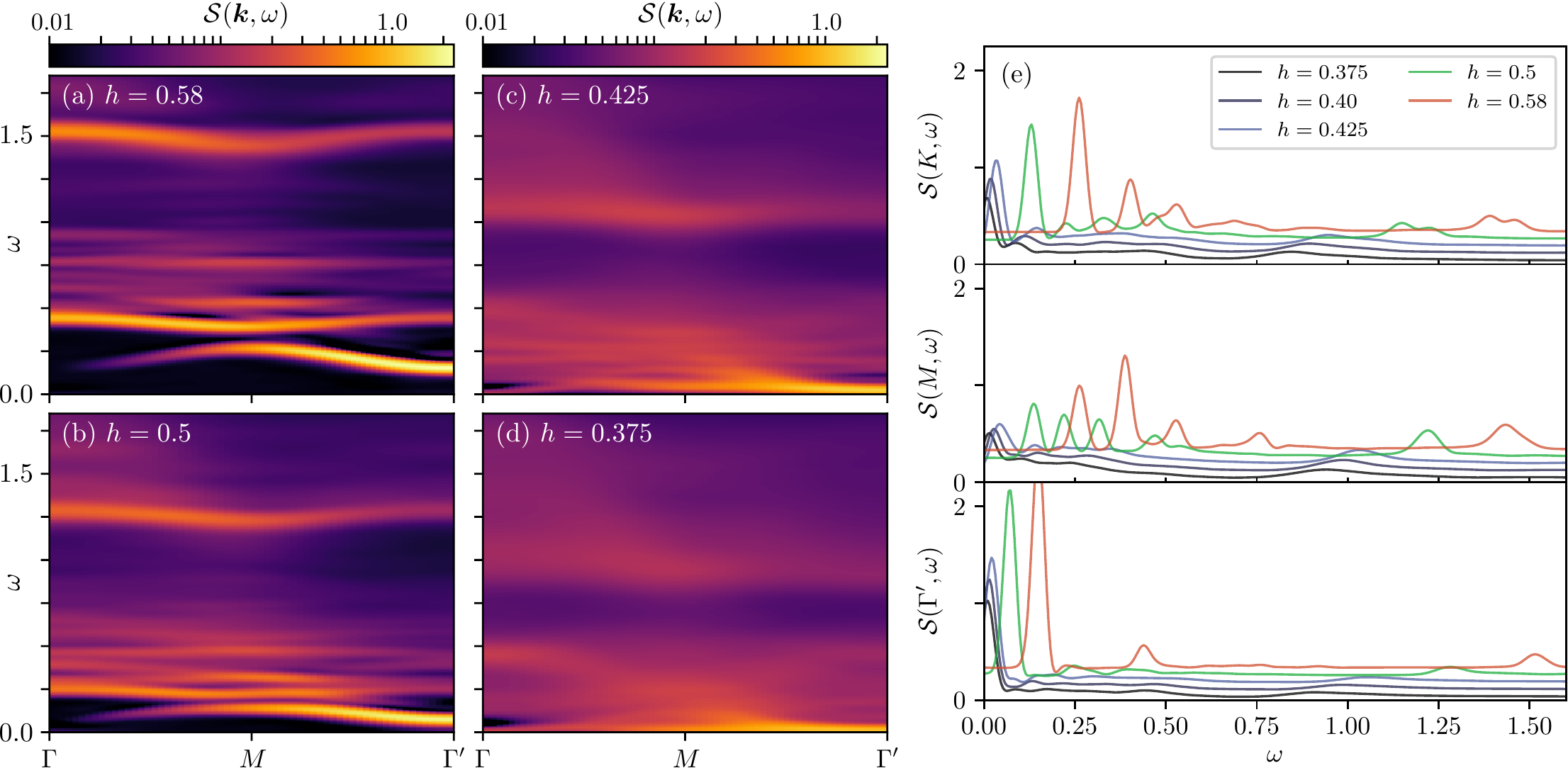}
        \caption{
            Dynamical spin-structure factor $\mathcal S(k,\omega)$ in the field-polarized phase
            along $\Gamma$--$M$--$\Gamma'$ at:
            (a) $h=0.58$, (b) $h=0.5$, (c) $h=0.425$, and (d) $h=0.375$.
            (e) $\mathcal S(k,\omega)$ at high-symmetry points $K$, $M$, and $\Gamma'$
            for different $h$.
            An vertical offset is used for better visibility.
            In all plots, $\mathcal S(k,\omega)$ is normalized as given in the main text.
        }
        \label{fig:dsf_pol}
    \end{figure*}
 
    The \emph{dynamical spin-structure factor} $\mathcal S(\bm{k},\omega)$ contains information about the excitation spectrum
    and is experimentally accessible via scattering experiments, in particular inelastic neutron scattering.
    We consider $\mathcal S(\bm k,\omega) = \sum_{\gamma=\{x,y,z\}} \mathcal S^{\gamma\gamma} (\bm k,\omega)$ 
    with $\mathcal S^{\gamma\gamma} (\bm k,\omega)$ being the spatio-temporal Fourier transform of the dynamical correlations 
    \begin{equation}
        \mathcal S^{\gamma\gamma} (\bm k,\omega) = N \int dt ~ e^{i \omega t} \sum_{a,b} e^{i (\bm r_b - \bm r_a) \cdot \bm k} ~ C^{\gamma\gamma}_{ab}(t) ~,
    \end{equation}
    where $\gamma = \{x,y,z\}$ is the spin component,
    $r_a$ and $r_b$ are the spatial positions of the spins,
    and diagonal elements $\mathcal S^{xx}$, $\mathcal S^{yy}$, and $\mathcal S^{zz}$ are considered.
    $N$ is defined by normalizing $\mathcal S^{\gamma\gamma} (\bm k,\omega)$ as 
    $\int d\omega \int d\bm k ~ \mathcal S^{\gamma\gamma} (\bm k,\omega) = \int d\bm k$.
    $C^{\gamma\gamma}_{ab} (t)$ denotes the dynamical spin-spin correlation
    \begin{align}
        C_{ab}^{\gamma\gamma}(t)    &= \langle \psi_0 | S_a^\gamma (t) S_b^\gamma (0) | \psi_0 \rangle \nonumber \\
                                    &= \langle \psi_0 | U(-t) S_a^\gamma U(t) S_b^\gamma | \psi_0 \rangle \nonumber \\
                                    &= \langle \psi_0 | S_a^\gamma U(t) S_b^\gamma | \psi_0 \rangle~, \label{eqn:Cij_t}
    \end{align}
    where the unitary time-evolution operator $U(t) = e^{-i(H-E_0)t}$ is modified
    by subtracting the ground state energy $E_0$.  
    Thus, the time-evolution $U(-t)$ acts trivially on the ground state $\langle \psi_0| U(-t) = \langle \psi_0 |$.
    Following Ref.~[\onlinecite{zaletel_time-evolving_2015}], we express $U(t)$
    into a \emph{matrix product operator} (MPO) with discretized time steps.
    
    Equation (\ref{eqn:Cij_t}) provides the numerical protocol we employ:
    (i) Obtain the ground state wave function $|\psi_0 \rangle$ using iDMRG
    and enlarge the iDMRG cell along the cylindrical axis to make room for the excitation to spread spatially,
    (ii) apply spin operator $S_i^\gamma$ at site $i$,
    (iii) time-evolve the MPS by $U(t)$,
    (iv) apply $S_j^\gamma$ at $j$,
    and (v) compute the overlap.

    On the technical side,
    we first compute the spatial Fourier transform of $C_{ab}^{\gamma\gamma}(t)$,
    extend the time-signal using linear predictive coding~\cite{white_spectral_2008},
    and multiply with a gaussian to suppress ringing due to the finite-time window.
    The extension of the time-signal allows for much wider finite-time windows
    keeping a significant part of the simulated real-time dynamics.
    All spectra shown in the remainder have a broadening of $\sigma_\omega=0.018$
    due to multiplying the real-time data with a Gaussian of width $\sigma_t=55.8$.
    The real-time data is obtained for times up to $T=120$ on cylinders with \emph{rhombic} geometry and $L_\text{circ}=6$.

    In the following, we discuss $\mathcal S(\bm k, \omega)$ within the topological phase and the polarized phase.
    Simulating the dynamics within the intermediate regime is left for future work
    as the necessary bond dimension for encoding the ground state is to large to achieve appreciably long times in the time-evolution. 

    \subsection{Topological Phase}

    Near $h=0$, see Fig. \ref{fig:dsf}(a),
    the numerically obtained $\mathcal S(\bm k, \omega)$ exhibits the features of the analytic solution\cite{knolle_dynamics_2014,knolle_dynamics_2015}
    with some adjustments due to the cylindrical geometry~\cite{gohlke_dynamics_2017}.
    Firstly, this involves a low-energy peak at $\omega\approx0.03$ %
    of which its spectral weight is shifted towards $\Gamma'$
    due to the antiferromagnetic nearest-neighbor spin-spin correlation
    caused by the antiferromagnetic Kitaev exchange. 
    When using a cylindrical geometry, an additional $\delta$-peak with finite spectral weight occurs at the two-flux energy.
    This $\delta$-peak, together with the finite-time evolution and subsequent broadening in frequency space,
    hides the two-flux gap.
    Nevertheless, the $\delta$-peak position coincides with the two-flux gap%
    \footnote{The two-flux gap is shifted towards smaller frequencies for narrow cylinders.},
    $\Delta_{2} \approx 0.03$.

    Secondly, a broad continuum exists, that is cut off near $\omega\approx1.5$.
    Increasing $h$ to $0.1$ and $0.2$, cf. Fig.~\ref{fig:dsf}(b) and (c), only leads to minor changes of the spectrum. 
    Most notably, the low-energy peak develops a shoulder towards slightly elevated energies,
    and the cut-off at $\omega\approx1.5$ is blurred out.
    Both features are more prominent in the line plots, Fig.~\ref{fig:dsf}(g).
    Any changes to the low-energy spectrum near or even below the two-flux gap are hidden in the energy resolution 
    caused by the finite time-evolution.

    In order to get a qualitative view on how the magnetic field affects the spectrum,
    we investigate the effect of both, $K_3$ and $h$.
    
    For $K_3=-0.25$ and $h=0.0$, Fig. \ref{fig:dsf}(d), the low-energy peak gets elevated to $\omega \approx 0.2$.
    This peak originates from a single fermion bound to a pair of fluxes\cite{knolle_dynamics_2015}
    and its shift is caused by $K_3$ increasing the two-flux gap.
    The fermion continuum starts at $\omega\approx0.4$,
    and the upper cut-off of the continuum remains near $\omega\approx1.5$.
    Both edges are sharp.
    Note that, $K_3 = -0.25$ has a similar correlation length as $h=0.2$
    as discussed above in relation to Fig.~\ref{fig:top_hp3_K3}.
    Yet, the corresponding spectra, Fig.~\ref{fig:dsf}(c) and (d), are qualitatively different,
    in that for $h=0.2$ the spectral weight is shifted significantly towards zero with no observable gap. 

    Upon increasing $h$ to $0.1$, the low-energy peak splits into at least three peaks, two of them develop a dispersion.
    Due to the field, the fluxes acquire a finite hopping amplitude and become mobile.
    The fluxes are thence no longer required to lie on neighboring plaquettes, but instead may separate.
    Hence, the mode describing a fermion bound to the two-flux pair generally attains more structure\cite{theveniaut_bound_2017}.
    Moreover, interaction between fluxes may induce further dispersion\cite{lahtinen_topological_2010,lahtinen_interacting_2011}.
    At further elevated fields, cf. Fig.~\ref{fig:dsf}(f) at $h=0.175$,
    somewhat before the phase transition into the intermediate regime%
    \footnote{In analogy to the findings in [\onlinecite{jiang_possible_2011}], the additional $K_3$ term shifts the critical field. For $K_3=-0.25$ the transition occurs at $h_{c1,AF}=0.19$.},
    the splitting increases with lots of the spectral weight shifting to the peak that is lowest in energy.
    The spectral gap reduces significantly with $h$ and has its minimum at the $\Gamma$ and $\Gamma'$ high-symmetry points.
    
    \subsection{Polarized Phase}
    
    From linear spin-wave theory (LSWT) it is known that the magnons are topological.
    Their bands carry a $\pm1$ Chern number and chiral edge modes have been observed on a slab geometry\cite{mcclarty_topological_2018,joshi_topological_2018}.
    But LSWT is only applicable for fields above the classical critical field $h_{\text{clas}}=1/\sqrt 3\approx0.58$.
    Here, we focus on the bulk excitation spectrum at fields between the numerically observed, $h_{c2,AF}\approx0.36$, and the classical critical field.
    Results for larger fields are presented in Ref. [\onlinecite{mcclarty_topological_2018}] using the same method.

    Beginning our discussion at the classical critical field $h=0.58$
    shown in Fig.~\ref{fig:dsf_pol}(a), we observe two magnon-bands
    with a minimum of $\omega\approx0.15$ at the high-symmetry points $\Gamma$ and $\Gamma'$.
    The two-magnon continuum has some overlap with the upper magnon band.
    With lowering the field, the magnon bands move down in energy 
    and flatten in the sense that their bandwidth decreases.
    At $h=0.5$, cf. Fig.~\ref{fig:dsf_pol}(b), the continuum already overlaps with major parts of the upper magnon band.
    This opens decay channels, limiting its lifetime, and consequently broadening the mode.

    Approaching the transition, cf. Fig.~\ref{fig:dsf_pol}(d) at $h=0.375$ and (c) at $h=0.425$,
    $\mathcal S(\omega,\bm k)$ shows a very broad continuum ranging down to almost zero energy,
    where also most of the spectral weight is observed.
    The upper magnon band is completely obscured by the multi-magnon continuum
    and lots of the spectral weight is distributed over a wide range in energy.
    The lower edge of the spectrum flattens towards the transition, 
    which is even more evident in the line plots shown in Fig.~\ref{fig:dsf_pol}(e).
    In particular at $h=0.375$ the low-energy peaks shift down to almost zero energy simultaneously
    at the $K$, $M$, and $\Gamma'$ high-symmetry points,
    with most of the spectral weight still appearing above the $\Gamma'$-point.
    
    This reproduces to some extent the phenomenology of LSWT,
    namely that the lower magnon band flattens completely while decreasing to zero energy\cite{janssen_honeycomb-lattice_2016,mcclarty_topological_2018},
    yet it occurs at lower fields than in LSWT.
    On the other hand, a clear remnant of the single magnon branch cannot be observed close to the transition 
    as it overlaps and merges with the multi-magnon continuum.    
    It may be possible that the single magnon branch is still dispersive, even though with a significantly reduced bandwidth.

    A feature in the spectrum not mentioned so far,
    emerges at around $\omega\approx1$ at magnetic fields near the transition.
    Initially this high-energy feature is very broad in energy,
    but sharpens and moves to higher energy upon increasing the field.
    At $h=0.5$ ($h=0.58$) it appears around $\omega\approx1.25$ ($\omega\approx1.5$) and exhibits a slight dispersion.
    At even larger fields, beyond what is presented here,
    the high-energy feature moves up in energy with a linear dependence on the field
    and twice the slope compared to the single-magnon excitations.
    Furthermore, the high-energy feature is situated at the upper edge of the two magnon continuum.
    Its intensity first increases, but starts to decrease at higher fields.
    It would be interesting to investigate,
    if this may be due to the appearance
    of an anti-bound state\cite{seibold_theory_2008} of two magnons 
    experiencing a repulsive interaction on account of the antiferromagnetic Kitaev
    exchange interaction between two adjacent flipped spins.

	\section{Conclusion}
	\label{scn:dis}
    We confirm the vastly different phenomenology between ferromagnetic and antiferromagnetic Kitaev interaction,
    if a magnetic field along $[111]$ direction is applied.
    In case of ferromagnetic Kitaev coupling, only a single magnetic transition is observed,
    that separates a low-$h$ topological phase from the large-$h$ field-polarized phase.
    Whereas for antiferromagnetic Kitaev coupling, the topological phase is more stable
    and an intermediate regime exists, that is possibly gapless.
    The topological order of the low-$h$ phase and its non-abelian anyonic excitations are verified
    by extracting the topological entanglement entropy.
    In addition to Ref.~[\onlinecite{jiang_possible_2011}],
    the topological order obtained with a finite three-spin term or when applying a weak magnetic field
    is the same also for antiferromagnetic Kitaev coupling.

    Upon applying the magnetic field,
    the spectral gap in the dynamical spin-structure factor remains within the frequency resolution
    and the overall spectrum exhibits only minor changes.    
    However, the dynamical spin-structure factor is remarkably different when applying the three-spin term
    lifting the spectral gap, both due to the flux gap increasing and the fermions gapping out.
    When a combination of magnetic field and three-spin term is applied,
    we observe a drastic reduction of the spectral gap with increased field
    and more structure in the low-energy peak corresponding to a bound state of a flux pair and a fermion.
    This additional structure is due to the fluxes becoming mobile and the flux-pair may separate
    providing a richer energy manifold for that bound state.
    Upon approaching the intermediate regime, the spectral gap reduces with a minimum at the $\Gamma'$ high-symmetry point.
    We can conclude, that even though the energy gap opens in a similar manner 
    when either the magnetic field or three-spin term is varied,
    the dynamical spin-structure factor exhibits a remarkably different low-energy structure. 
    Thus, additional terms in perturbation theory, other then the three-spin term preserving integrability,
    are relevant to describe the dynamical spin-structure factor in the topological phase.

    When approaching the intermediate region from high-fields,
    we observe a strong reduction in frequency with a simultaneous flattening of the lower magnon band.
    A broad continuum develops, that ranges down to the lowest frequencies and merges with the single magnon branch.
    It remains an open question, 
    whether this flattening could be attributed to the collapse of the lower magnon branch, as observed within LSWT, 
    or rather to multi-magnon excitations obscuring any dispersion of the very same magnon branch.
    Nonetheless, the flat gap closing as such is interesting in various aspects
    as it may indicate exotic spin states like a quantum spin liquid.

    \section{Acknowledgements} 
    
    We are grateful to Bal\'asz D\'ora, Lucas Janssen, Paul McClarty, Karlo Penc, Jeffrey G. Rau and Ruben Verresen for stimulating discussions.
    This work was supported in part by DFG via SFB 1143. FP acknowledges the support of the DFG Research Unit FOR 1807 through grants no. PO 1370/2- 1, TRR80, and the Nanosystems Initiative Munich (NIM) by the German Excellence Initiative. 

	\appendix
    
    \section{Finite-size dependents within intermediate phase}
    \label{app:fs_int}

    \begin{figure}[tb]
		\includegraphics[width=\linewidth]{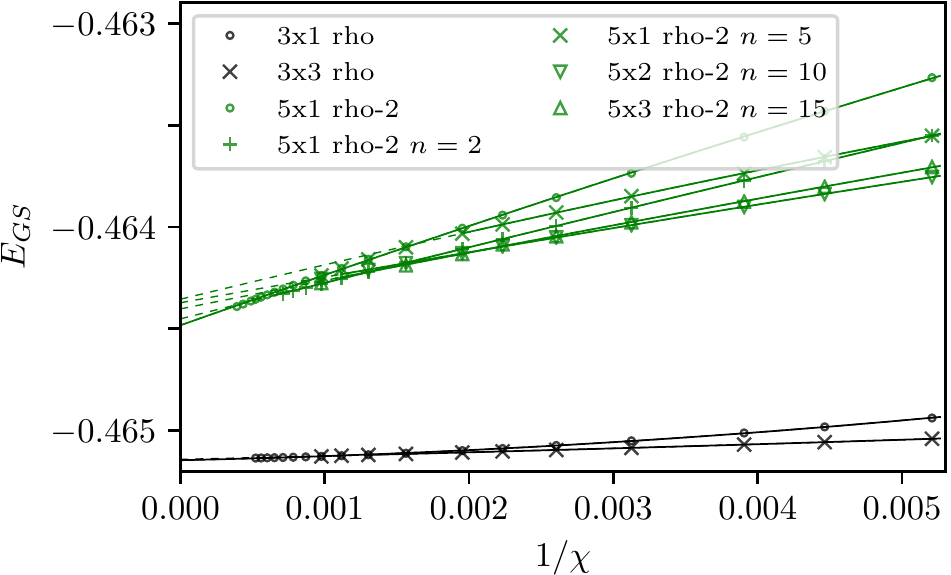}
		\caption{Comparison of the ground state energy $E_{GS}$ vs bond-dimension $\chi$ for different geometries and sizes of the iDMRG cluster at a single field strength of $h=0.275$.
        For small $\chi$, larger iDMRG cluster have a smaller $E_{GS}$.
        However, in the limit $1/\chi \rightarrow 0$, $E_{GS}$ is of very similar value for all geometries used.
        In fact, large iDMRG cluster show a phase transition from an ordered ground state at small $\chi$ to a translational invariant ground state at large $\chi$ captured by the smallest iDMRG cluster.}
        \label{fig:h055_E_invchi}
	\end{figure}	
    
    Here, we investigate the intermediate regime with respect to possible finite-size effects as well as finite bond dimension of the matrix product state (MPS).
    Figure \ref{fig:h055_E_invchi} provides a comparison of the ground state energy $E_{GS}$ for two different geometries, \emph{rhombic} with $L_\text{circ}=6$ or \emph{rhombic-2} with $L_\text{circ}=10$, as well as several different sizes of the iDMRG cluster at a magnetic field of $h=0.275$.
    Similar checks are done at different $h$.

    In case of \emph{rhombic-2} with $L_\text{circ} = 10$ (green symbols), the smallest cluster is similar to a single fundamental unit cell with two sites (green circles), that is repeated along a chain winding around the cylinder. 
    Next larger clusters are: four sites ($n=2$ fundamental unit cells, green 'x'), 10 sites ($n=5$, green '+'), 20 sites ($n=10$, green lower triangle), and 30 sites ($n=15$, green upper triangle).
    When using small bond-dimensions $\chi < 500$, larger iDMRG clusters result in lower ground state energies $E_{GS}$.
    Upon increasing $\chi$, the different energies approach each other until eventually a transition to the ground state of a smaller cluster occurs, e.g., at $\chi \ge 512$ the 10 site cluster ('x') has the same ground state properties as the fundamental unit cell (circles).
    Such a \emph{$\chi$-transition} is unphysical and a mere property of truncating the MPS.
    
    In case of \emph{rhombic} with $L_\text{circ} = 6$ (black symbols) in Fig. \ref{fig:h055_E_invchi}, the smallest iDMRG cluster is a single ring with three fundamental unit cell along the circumference ('3x1', black circles).
    Larger clusters of three repetitions along the cylinder ('3x3', black triangles) and six repetitions (not shown, but equivalent to '3x3') are checked.
    As above, a similar $\chi$-transition at $\chi \approx 800$ is found, where for smaller $\chi$ the '3x3' has a lower $E_{GS}$, but transitions to the same state as '3x1' for larger $\chi$.
    
    In conclusion, the ground states for larger $\chi$ are not exhibiting any broken translational symmetry and may resemble the physical ground state.
    Thus, the use of iDMRG cell composed of a single fundamental unit cell is justified for computing the phase diagram shown in Fig. \ref{fig:pd}. 

    \begin{figure}[tb]
		\includegraphics[width=\linewidth]{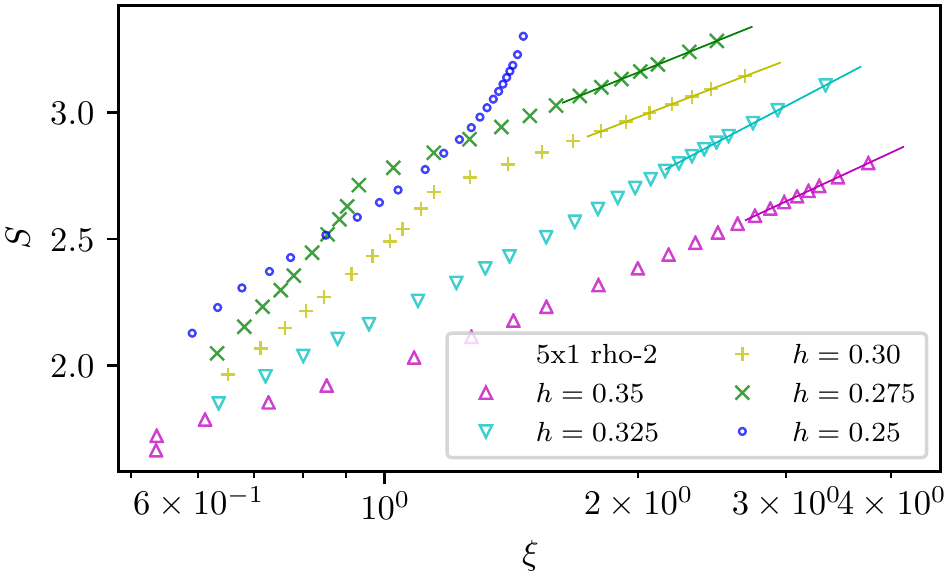}
        \caption{
        	Entanglement entropy $S$ of a bipartition of the cylinder over correlation length $\xi$ for various bond dimension $\chi$ to check for possible finite-$\chi$ scaling.
        	Data is shown for different magnetic field strength $h = 0.25, 0.275, 0.3, 0.325, 0.35$ across the intermediate regime.
        	Lines are fits to the five points with largest $\xi$ at each field.
        	}
        \label{fig:itm_xi_Sbip}
	\end{figure}

    The previous results signify, that large bond dimensions are necessary to resemble the physical ground state.
    We cannot say for sure, that the $\chi$ we are able to achieve are already sufficient, thus any statement regarding the intermediate region has to be taken with care.
    Nonetheless, let us assume the MPS do reflect physical properties of the underlying phase and apply a finite-$\chi$ scaling.
    For $h=0.275,0.3,0.325$, and $0.35$ we obtain a $S_{E,\chi} = c/6 \log \xi_\chi + \text{const}$ scaling typical for a gapless phase~\cite{calabrese_entanglement_2004,tagliacozzo_scaling_2008}, see Fig. \ref{fig:itm_xi_Sbip}.
    Linear regression of the five points with largest $\chi$ reveal slopes corresponding to central charges of $c=3.49$ at $h=0.275$, $c=3.31$ at $h=0.30$, $c=4.54$ at $h=0.325$, and $c=4.01$ for $h=0.35$, all for \emph{rhombic-2} with $L_\text{circ}=10$.
    We want to remark, that not all of the $c$ represent \emph{physical} central charges.
    $h = 0.3$ has a very similar behaviour in terms of $S_E$ vs $\xi$, but a slightly smaller $c$, which may converge to $3.5$ for larger $\chi$.
    For $h = 0.25$, $\chi$ does not yet suffice to enter a linear $S_E \propto \log \xi$ regime.  

	\bibliography{KH111.bib}
	
\end{document}